# Solving Big Data Challenges for Enterprise Application Performance Management

Tilmann Rabl
Middleware Systems
Research Group
University of Toronto, Canada
tilmann@msrg.utoronto.ca

Mohammad Sadoghi
Middleware Systems
Research Group
University of Toronto, Canada
mo@msrg.utoronto.ca

Hans-Arno Jacobsen
Middleware Systems
Research Group
University of Toronto, Canada
arno@msrg.utoronto.ca

Sergio Gómez-Villamor
DAMA-UPC
Universitat Politècnica de
Catalunya, Spain
sgomez@ac.upc.edu

Victor Muntés-Mulero
CA Labs Europe
Barcelona, Spain
victor.muntes@ca.com

Serge Mankovskii
CA Labs
San Francisco, USA
serge.mankovskii@ca.com

## ABSTRACT

As the complexity of enterprise systems increases, the need for monitoring and analyzing such systems also grows. A number of companies have built sophisticated monitoring tools that go far beyond simple resource utilization reports. For example, based on instrumentation and specialized APIs, it is now possible to monitor single method invocations and trace individual transactions across geographically distributed systems. This high-level of detail enables more precise forms of analysis and prediction but comes at the price of high data rates (i.e., big data). To maximize the benefit of data monitoring, the data has to be stored for an extended period of time for ulterior analysis. This new wave of big data analytics imposes new challenges especially for the application performance monitoring systems. The monitoring data has to be stored in a system that can sustain the high data rates and at the same time enable an up-to-date view of the underlying infrastructure. With the advent of modern key-value stores, a variety of data storage systems have emerged that are built with a focus on scalability and high data rates as predominant in this monitoring use case.

In this work, we present our experience and a comprehensive performance evaluation of six modern (open-source) data stores in the context of application performance monitoring as part of CA Technologies initiative. We evaluated these systems with data and workloads that can be found in application performance monitoring, as well as, on-line advertisement, power monitoring, and many other use cases. We present our insights not only as performance results but also as lessons learned and our experience relating to the setup and configuration complexity of these data stores in an industry setting.

## 1. INTRODUCTION

Large scale enterprise systems today can comprise complete data centers with thousands of servers. These systems are heterogeneous and have many interdependencies which makes their administration a very complex task. To give administrators an on-line view of the system health, monitoring frameworks have been developed. Common examples are Ganglia [20] and Nagios [12]. These are widely used in open-source projects and academia (e.g., Wikipedia[1]). However, in industry settings, in presence of stringent response time and availability requirements, a more thorough view of the monitored system is needed. Application Performance Management (APM) tools, such as Dynatrace[2], Quest PerformaSure[3], AppDynamics[4], and CA APM[5] provide a more sophisticated view on the monitored system. These tools instrument the applications to retrieve information about the response times of specific services or combinations of services, as well as about failure rates, resource utilization, etc. Different monitoring targets such as the response time of a specific servlet or the CPU utilization of a host are usually referred to as *metrics*. In modern enterprise systems it is not uncommon to have thousands of different metrics that are reported from a single host machine. In order to allow for detailed on-line as well as off-line analysis of this data, it is persisted at a centralized store. With the continuous growth of enterprise systems, sometimes extending over multiple data centers, and the need to track and report more detailed information, that has to be stored for longer periods of time, a centralized storage philosophy is no longer viable. This is critical since monitoring systems are required to introduce a low overhead – i.e., 1-2% on the system resources [2] – to not degrade the monitored system's performance and to keep maintenance budgets low. Because of these requirements, emerging storage systems have to be explored in order to develop an APM platform for monitoring big data with a tight resource budget and fast response time.

---

[1]Wikipedia's Ganglia installation can be accessed at http://ganglia.wikimedia.org/latest/.
[2]Dynatrace homepage - http://www.dynatrace.com
[3]PerformaSure homepage - http://www.quest.com/performasure/
[4]AppDynamics homepage - http://www.appdynamics.com
[5]CA APM homepage - http://www.ca.com/us/application-management.aspx

APM has similar requirements to current Web-based information systems such as weaker consistency requirements, geographical distribution, and asynchronous processing. Furthermore, the amount of data generated by monitoring applications can be enormous. Consider a common customer scenario: *The customer's data center has 10K nodes, in which each node can report up to 50K metrics with an average of 10K metrics.* As mentioned above, the high number of metrics result from the need for a high-degree of detail in monitoring, – an individual metric for response time, failure rate, resource utilization, etc. of each system component can be reported. In the example above, with a modest monitoring interval of 10 seconds, 10 million individual measurements are reported per second. Even though a single measurement is small in size, below 100 bytes, the mass of measurements poses similar big data challenges as those found in Web information system applications such as on-line advertisement [9] or on-line analytics for social Web data [25]. These applications use modern storage systems with focus on scalability as opposed to relational database systems with a strong focus on consistency. Because of the similarity of APM storage requirements to the requirements of Web information system applications, obvious candidates for new APM storage systems are key-value stores and their derivatives. Therefore, we present a performance evaluation of different key-value stores and related systems for APM storage.

Specifically, we present our benchmarking effort on open source key-value stores and their close competitors. We compare the throughput of Apache Cassandra, Apache HBase, Project Voldemort, Redis, VoltDB, and a MySQL Cluster. Although, there would have been other candidates for the performance comparison, these systems cover a broad area of modern storage architectures. In contrast to previous work (e.g., [7, 23, 22]), we present details on the maximum sustainable throughput of each system. We test the systems in two different hardware setups: (1) a memory- and (2) a disk-bound setup.

Our contributions are threefold: (1) we present the use case and *big data challenge* of application performance management and specify its data and workload requirements; (2) we present an up-to-date performance comparison of six different data store architectures on two differently structured compute clusters; and, (3) finally, we report on details of our experiences with these systems from an industry perspective.

The rest of the paper is organized as follows. In the next section, we describe our use case: application performance management. Section 3 gives an overview of the benchmarking setup and the testbed. In Section 4, we introduce the benchmarked data stores. In Section 5, we discuss our benchmarking results in detail. Section 6 summarizes additional findings that we made during our benchmarking effort. In Section 7, we discuss related work before concluding in Section 8 with future work.

## 2. APPLICATION PERFORMANCE MANAGEMENT

Usually enterprise systems are highly distributed and heterogeneous. They comprise a multitude of applications that are often interrelated. An example of such a system can be seen in Figure 1. Clients connect to a frontend, which can be a Web server or a client application. A single client interaction may start a transaction that can span over more than a thousand components, which can be hosted on an equal number of physical machines [26]. Nevertheless, response time is critical in most situations. For example, for Web page loads the consumer expectation is constantly decreasing and is already as low as 50 ms to 2 s [3]. In a highly distributed system, it is difficult to determine the root cause of performance deterioration especially since it is often not tied to a single component, but to a specific interaction of components. System components themselves are highly heterogeneous due to the constant changes in application software and hardware. There is no unified code base and often access to the entire source code is not possible. Thus, an in depth analysis of the components or the integration of a profiling infrastructure is not possible.

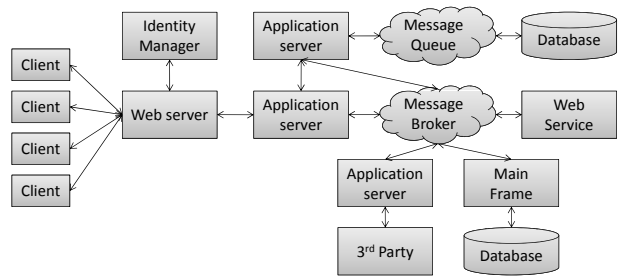

**Figure 1: Example of an enterprise system architecture**

To overcome this challenges, application performance management systems (APM) have been developed and are now a highly profitable niche in enterprise system deployment. APM refers to the monitoring and managing of enterprise software systems. There are two common approaches to monitor enterprise systems: (1) an API-based approach, which provides a programming interface and a library that has to be utilized by all monitored components; (2) a black-box approach, which instruments the underlying system components or virtual machines to obtain information about the monitored system. The first approach gives a high degree of freedom to the programmer on how to utilize the monitoring toolbox. A popular example is the ARM standard [1]. In this approach every component has to implement the ARM API that is available for C and Java. Prominent ARM-instrumented applications are the Apache HTTP server and IBM DB2. Although several common enterprise software systems are already ARM enabled, it is often not feasible to implement the ARM API in legacy systems. In general, this solution is often not possible, especially when $3^{rd}$ party software components are used. The instrumentation of virtual machines and system libraries is a non-intrusive way of monitoring an application. In the case of Java programs, this is enabled by the Virtual Machine Tool Interface that was specified in JSR-163 and introduced in J2SE 5.0 [11]. The intention of JSR-163 is to present an interface for profiling and debugging. Byte code instrumentation allows to augment software components with agents that have access to the state and the method invocations. This approach enables monitoring components, tracing transactions, and performing root cause analysis without changing the code base of the monitored system. Another benefit of this approach is the low performance overhead incurred.

By making it possible to capture every method invocation in a large enterprise system, APM tools can generate a vast amount of data. However, in general, only specific method invocations are actually of interest. Most notably, these are communication methods such as RMI calls, Web service calls, socket connections and such. Still, current systems process thousands of transactions per second with each transaction being processed by multiple nodes. Due to the resulting amount of data, monitoring agents do not report every single event but instead aggregate events in fixed time intervals in the order of seconds. Nevertheless, each agent can report thousands of different measurements in each interval. In larger deployments, i.e., hundreds to thousands of hosts, this results in a sustained rate of millions of measurements per second. This information is valuable for later analysis and, therefore, should be stored in a long-term archive. At the same time, the most recent data has to be readily available for on-line monitoring and for generating emergency notifications. Typical requirements are sliding window aggregates over the most recent data of a certain type of measurement or *metric* as well as aggregates over multiple metrics of the same type measured on different machines. For instance two typical on-line queries are:

- What was the maximum number of connections on host $X$ within the last 10 minutes?
- What was the average CPU utilization of Web servers of type $Y$ within the last 15 minutes?

For the archived data more analytical queries are as follows:

- What was the average total response time for Web requests served by replications of servlet $X$ in December 2011?
- What was maximum average response time of calls from application $Y$ to database $Z$ within the last month?

While the on-line queries have to be processed in real-time, i.e., in subsecond ranges, historical queries may finish in the order of minutes. In comparison to the insertion rate, these queries are however issued rarely. Even large clusters are monitored by a modest number of administrators which makes the ad-hoc query rate rather small. Some of the metrics are monitored by certain triggers that issue notifications in extreme cases. However, the overall write to read ratio is 100:1 or more (i.e., write-dominated workloads). While the writes are simple inserts, the reads often scan a small set of records. For example, for a ten minute scan window with 10 seconds resolution, the number of scanned values is 60.

An important prerequisite for APM is that the performance of the monitored application should not be deteriorated by the monitoring (i.e., monitoring should not effect SLAs.) As a rule of thumb, a maximum tolerable overhead is five percent, but a smaller rate is preferable. This is also true for the size of the storage system. This means that for an enterprise system of hundreds of nodes only tens of nodes may be dedicated to archiving monitoring data.

## 3. BENCHMARK AND SETUP

As explained above, the APM data is in general relatively simple. It usually consists of a metric name, a value, and a time stamp. Since many agents report their data in fixed length intervals, the data has the potential to be aggregated over a (short) time-based sliding window, and may also contain additional aggregation values such as minimum, maximum, average, and the duration. A typical example could look as shown in Figure 2. The record structure is usually fixed.

As for the workload, the agents report the measurements for each metric in certain periodic intervals. These intervals are in the order of seconds. Using this approach, the data rate at the agent is in general constant regardless of the system load. Nevertheless, current APM tools make it possible to define different monitoring levels, e.g., basic monitoring mode, transaction trace mode, and incident triage mode, that results in different data rates. It is important to mention that the monitoring data is append only. Every new reported measurement is appended to the existing monitoring information rather than updating or replacing it. Since the agents report changes in the system in an aggregated manner, for example, every

**Table 1: Workload specifications**

| Workload | % Read | % Scans | % Inserts |
|---|---|---|---|
| R | 95 | 0 | 5 |
| RW | 50 | 0 | 50 |
| W | 1 | 0 | 99 |
| RS | 47 | 47 | 6 |
| RSW | 25 | 25 | 50 |

10 seconds, the queries on the reported measurements do not have latency as low as that found in OLTP use cases, rather a latency in the same order as the reporting interval is still adequate. As far as the storage system is concerned, the queries can be distinguished into two major types: (1) single value lookups to retrieve the most current value and (2) small scans for retrieving system health information and for computing aggregates over time windows.

Based on the properties of the APM use case described above, we designed a benchmark that models this use case, which at the same time is generic enough to be valid for similar monitoring applications. We used the popular Yahoo! Cloud Serving Benchmark (YCSB) benchmark suite [7] as a basis. YCSB is an extensible and generic framework for the evaluation of key-value stores. It allows to generate synthetic workloads which are defined as a configurable distribution of CRUD (create, read, update and delete) operations on a set of records. Records have a predefined number of fields and are logically indexed by a key. This generic data model can easily be mapped to a specific key-value or column-based data model. The reason for YCSB's popularity is that it comprises a data generator, a workload generator, as well as drivers for several key-value stores, some of which are also used in this evaluation.

Our data set consists of records with a single alphanumeric key with a length of 25 bytes and 5 value fields each with 10 bytes. Thus, a single record has a raw size of 75 bytes. This is consistent with the real data structure as shown in Figure 2.

We defined five different workloads. They are shown in Table 1. As mentioned above, APM data is append only – which is why we only included *insert*, *read*, and *scan* operations. Since not all tested stores support scans, we defined workloads with (RS,RSW) and without scans (R,RW,W). As explained above, APM systems exhibit a write to read ratio of 100:1 or more as defined in workloads However, to give a more complete view on the systems under test, we defined workloads that vary the write to read ratio. Workload R and RS are read-intensive where 50% of the read accesses in RS are scans. Workload RW and RSW have an equal ratio of reads and writes. These workloads are commonly considered write-heavy in other environments [7]. All access patterns were uniformly distributed. We also tested a write intensive workload with scans, but we omit it here due to space constraints.

We used two independent clusters for our tests: memory-bound cluster (Cluster M) and disk-bound cluster (Cluster D). Cluster M consists of 16 Linux nodes. Each node has two Intel Xeon quad core CPUs, 16 GB of RAM, and two 74 GB disks configured in RAID 0, resulting in 148 GB of disk space per node. The nodes are connected with a gigabit ethernet network over a single switch. Additionally, we used an additional server with the same configuration but an additional 6 disk RAID 5 array with 500 GB disk per node for a total of 2.5 TB of disk space. This RAID disk is mounted on all nodes and is used to store the binaries and configuration files. During the test, the nodes do, however, use only their local disks. Cluster D consists of a 24 Linux nodes, in which each node has two Intel Xeon dual core CPUs, 4 GB of RAM and a single 74 GB disk. The nodes are connected with a gigabit ethernet network over a sin-gle switch. We use the Cluster M for memory-bound experiments and the Cluster D for disk-bound tests.

| Metric Name | Value | Min | Max | Timestamp | Duration |
|---|---|---|---|---|---|
| HostA/AgentX/ServletB/AverageResponseTime | 4 | 1 | 6 | 1332988833 | 15 |

Figure 2: Example of an APM measurement

For Cluster M the size of the data set was set to 10 million records per node resulting 700 MB of raw data per node. The raw data in this context does not include the keys which increases the total data footprint. For Cluster D we tested a single setup with 150 million records for a total of 10.5 GB of raw data and thus making memory-only processing impossible. Each test was run for 600 seconds and the reported results are the average of at least 3 independent executions. We automated the benchmarking process as much as possible to be able to experiment with many different settings. For each system, we wrote a set of scripts that performed the complete benchmark for a given configuration. The scripts installed the systems from scratch for each workload on the required number of nodes, thus making sure that there was no interference between different setups. We made extensive use of the Parallel Distributed Shell (pdsh). Many configuration parameters were adapted within the scripts using the stream editor sed.

Our workloads were generated using 128 connections per server node, i.e., 8 connections per core in Cluster M. In Cluster D, we reduced the number of connection to 2 per core to not overload the system. The number of connections is equal to the number of independently simulated clients for most systems. Thus, we scaled the number of threads from 128 for one node up to 1536 for 12 nodes, all of them working as intensively as possible. To be on the safe side, we used up to 5 nodes to generate the workload in order to fully saturate the storage systems. So no client node was running more than 307 threads. We set a scan-length of 50 records as well as fetched all the fields of the record for read operations.

## 4. BENCHMARKED KEY-VALUE STORES

We benchmarked six different open-source key-value stores. We chose them to get an overview of the performance impact of different storage architectures and design decisions. Our goal was not only to get a pure performance comparison but also a broad overview of available solutions. According to Cartell, new generation data stores can be classified in four main categories [4]. We chose each two of the following classes according to this classification:

**Key-value stores:** Project Voldemort and Redis

**Extensible record stores:** HBase and Cassandra

**Scalable relational stores:** MySQL Cluster and VoltDB

Our choice of systems was based on previously reported performance results, popularity and maturity. Cartell also describes a fourth type of store, document stores. However, in our initial research we did not find any document store that seemed to match our requirements and therefore did not include them in the comparison. In the following, we give a basic overview on the benchmarked systems focusing on the differences. Detailed descriptions can be found in the referenced literature.

### 4.1 HBase

HBase [28] is an open source, distributed, column-oriented database system based on Google's BigTable [5]. HBase is written in Java, runs on top of Apache Hadoop and Apache ZooKeeper [15] and uses the Hadoop Distributed Filesystem (HDFS) [27] (also an open source implementation of Google's file system GFS [13]) in order to provide fault-tolerance and replication.

Specifically, it provides linear and modular scalability, strictly consistent data access, automatic and configurable sharding of data. Tables in HBase can be accessed through an API as well as serve as the input and output for MapReduce jobs run in Hadoop. In short, applications store data into tables which consist of rows and column families containing columns. Moreover, each row may have a different set of columns. Furthermore, all columns are indexed with a user-provided key column and are grouped into column families. Also, all table cells – the intersection of row and column coordinates – are versioned and their content is an uninterpreted array of bytes.

For our benchmarks, we used HBase v0.90.4 running on top of Hadoop v0.20.205.0. The configuration was done using a dedicated node for the running master processes (NameNode and SecondaryNameNode), therefore for all the benchmarks the specified number of servers correspond to nodes running slave processes (DataNodes and TaskTrackers) as well as HBase's region server processes. We used the already implemented HBase YCSB client, which required one table for all the data, storing each field into a different column.

### 4.2 Cassandra

Apache Cassandra is a second generation distributed key value store developed at Facebook [19]. It was designed to handle very large amounts of data spread out across many commodity servers while providing a highly available service without single point of failure allowing replication even across multiple data centers as well as for choosing between synchronous or asynchronous replication for each update. Also, its elasticity allows read and write throughput, both increasing linearly as new machines are added, with no downtime or interruption to applications. In short, its architecture is a mixture of Google's BigTable [5] and Amazon's Dynamo [8]. As in Amazon's Dynamo, every node in the cluster has the same role, so there is no single point of failure as there is in the case of HBase. The data model provides a structured key-value store where columns are added only to specified keys, so different keys can have different number of columns in any given family as in HBase. The main differences between Cassandra and HBase are columns that can be grouped into column families in a nested way and consistency requirements that can be specified at query time. Moreover, whereas Cassandra is a write-oriented system, HBase was designed to get high performance for intensive read workloads.

For our benchmark, we used the recent 1.0.0-rc2 version and used mainly the default configuration. Since we aim for a high write throughput and have only small scans, we used the default RandomPartitioner that distributes the data across the nodes randomly. We used the already implemented Cassandra YCSB client which required to set just one column family to store all the fields, each of them corresponding to a column.

### 4.3 Voldemort

Project Voldemort [29] is a distributed key-value store (developed by LinkedIn) that provides highly scalable storage system.

With a simpler design compared to a relational database, Voldemort neither tries to support general relational model nor to guarantee full ACID properties, instead it simply offers a distributed, fault-tolerant, persistent hash table. In Voldemort, data is automatically replicated and partitioned across nodes such that each node is responsible for only a subset of data independent from all other nodes. This data model eliminates the central point of failure or the need for central coordination and allows cluster expansion without rebalancing all data, which ultimately allow horizontal scaling of Voldemort.

Through simple API, data placement and replication can easily be tuned to accommodate a wide range of application domains. For instance, to add persistence, Voldemort can use different storage systems such as embedded databases (e.g., BerkeleyDB) or stand-alone relational data stores (e.g., MySQL). Other notable features of Voldemort are in-memory caching coupled with storage system – so a separate caching tier is no longer required and multi-version data model for improved data availability in case of system failure.

In our benchmark, we used version 0.90.1. with the embedded BerkeleyDB storage and the already implemented Voldemort YCSB client. Specifically, when configuring the cluster, we set two partitions per node. In both clusters, we set Voldemort to use about 75% of the memory whereas the remaining 25% was used for the embedded BerkeleyDB storage. The data is stored in a single table where each key is associated with an indexed set of values.

### 4.4 Redis

Redis [24] is an in-memory, key-value data store with the data durability option. Redis data model supports strings, hashes, lists, sets, and sorted sets. Although Redis is designed for in-memory data, depending on the use case, data can be (semi-) persisted either by taking snapshot of the data and dumping it on disk periodically or by maintaining an append-only log of all operations.

Furthermore, Redis can be replicated using a master-slave architecture. Specifically, Redis supports relaxed form of master-slave replication, in which data from any master can be replicated to any number of slaves while a slave may acts as a master to other slaves allowing Redis to model a single-rooted replication tree.

Moreover, Redis replication is non-blocking on both the master and slave, which means that the master can continue serving queries when one or more slaves are synchronizing and slaves can answer queries using the old version of the data during the synchronization. This replication model allows for having multiple slaves to answer read-only queries resulting in highly scalable architecture.

For our benchmark, we used version 2.4.2. Although a Redis cluster version is expected in the future, at the time of writing this paper, the cluster version is in an unstable state and we were not able to run a complete test. Therefore, we deployed a single-node version on each of the nodes and used the Jedis[6] library to implement a distributed store. We updated the default Redis YCSB client to use ShardedJedisPool, a class which automatically shards and accordingly accesses the data in a set of independent Redis servers. This gives considerable advantage to Redis in our benchmark since there is no interaction between the Redis instances. For the storage of the data, YCSB uses a hash map as well as a sorted set.

### 4.5 VoltDB

VoltDB [30] is an ACID compliant relational in-memory database system derived from the research prototype H-Store [17]. It has a shared nothing architecture and is designed to run on a multi-node cluster by dividing the database into disjoint partitions by making each node the unique owner and responsible for a subset of the partitions. The unit of transaction is a stored procedure which is Java interspersed with SQL. Forcing stored procedures as the unit of transaction and executing them at the partition containing the necessary data makes it possible to eliminate round trip messaging between SQL statements. The statements are executed serially and in a single threaded manner without any locking or latching. The data is in-memory, hence, if it is local to a node a stored procedure can execute without any I/O or network access, providing very high throughput for transactional workloads. Furthermore, VoltDB supports multi-partition transactions, which require data from more than one partition and are therefore more expensive to execute. Multi-partition transactions can completely be avoided if the database is cleanly partitionable.

---

[6]Jedis, a Java client for Redis - http://github.com/xetorthio/jedis.

For our benchmarking we used VoltDB v2.1.3 and the default configuration. We set 6 sites per host which is the recommendation for our platform. We implemented an YCSB client driver for VoltDB that connects to all servers as suggested in the documentation. We set a single table with 5 columns for each of the fields and the key as the primary key as well as being the column which allows VoltDB for computing the partition of the table. This way, as read, write and insert operations are performed for a single key, they are implemented as single-partition transactions and just the scan operation is a multi-partition transaction. Also, we implemented the required stored procedures for each of the operations as well as the VoltDB YCSB client.

### 4.6 MySQL

MySQL [21] is the world's most used relational database system with full SQL support and ACID properties. MySQL supports two main storage engines: MyISAM (for managing non-transactional tables) and InnoDB (for providing standard transactional support). In addition, MySQL delivers an in-memory storage abstraction for temporary or non-persistent data. Furthermore, the MySQL cluster edition is a distributed, multi-master database with no single point of failure. In MySQL cluster, tables are automatically sharded across a pool of low-cost commodity nodes, enabling the database to scale horizontally to serve read and write-intensive workloads.

For our benchmarking we used MySQL v5.5.17 and InnoDB as the storage engine. Although MySQL cluster already provides shared-nothing distribution capabilities, instead we spread independent single-node servers on each node. Thus, we were able to use the already implemented RDBMS YCSB client which connects to the databases using JDBC and shards the data using a consistent hashing algorithm. For the storage of the data, a single table with a column for each value was used.

## 5. EXPERIMENTAL RESULTS

In this section, we report the results of our benchmarking efforts. For each workload, we present the throughput and the latencies of operations. Since there are huge variations in the latencies, we present our results using logarithmic scale. We will first report our results and point out significant values, then we will discuss the results. Most of our experiments were conducted on Cluster M which is the faster system with more main memory. Unless we explicitly specify the systems were tested on Cluster M.

### 5.1 Workload R

The first workload was Workload R, which was the most read intensive with 95% reads and only 5% writes. This kind of workload is common in many social web information systems where the read operations are dominant. As explained above, we used 10 million records per node, thus, scaling the problem size with the cluster size. For each run, we used a freshly installed system and loaded the data. We ran the workload for 10 minutes with maximum throughput. Figure 3 shows the maximum throughput for workload R for all six systems.

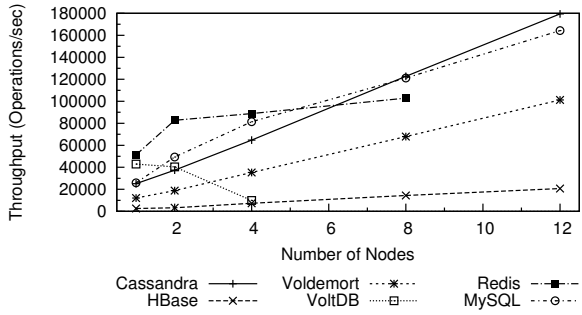

Figure 3: Throughput for Workload R

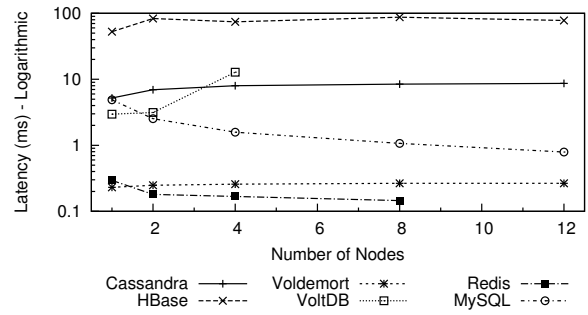

Figure 4: Read latency for Workload R

In the experiment with only one node, Redis has the highest throughput (more than 50K ops/sec) followed by VoltDB. There are no significant differences between the throughput of Cassandra and MySQL, which is about half that of Redis (25K ops/sec). Voldemort is 2 times slower than Cassandra (with 12K ops/sec). The slowest system in this test on a single node is HBase with 2.5K operation per second. However, it is interesting to observe that the three web data stores that were explicitly built for scalability in web scale – i.e. Cassandra, Voldemort, and HBase – demonstrate a nice linear behavior in the maximum throughput.

As discussed previously, we were not able to run the cluster version of Redis, therefore, we used the Jedis library that shards the data on standalone instances for multiple nodes. In theory, this is a big advantage for Redis, since it does not have to deal with propagating data and such. This also puts much more load on the client, therefore, we had to double the number of machines for the YCSB clients for Redis to fully saturate the standalone instances. However, the results do not show the expected scalability. During the tests, we noticed that the data distribution is unbalanced. This actually caused one Redis node to consistently run out of memory in the 12 node configuration[7]. For VoltDB, all configurations that we tested showed a slow-down for multiple nodes. It seems that the synchronous querying in YCSB is not suitable for a distributed VoltDB configuration. For MySQL we used a similar approach as for Redis. Each MySQL node was independent and the client managed the sharding. Interestingly, the YCSB client for MySQL did a much better sharding than the Jedis library, and we observed an almost perfect speed-up from one to two nodes. For higher number of nodes the increase of the throughput decreased slightly but was comparable to the throughput of Cassandra.

Workload R was read-intensive and modeled after the requirements of web information systems. Thus, we expected a low latency for read operations at the three web data stores. The average latencies for read operations for Workload R can be seen in Figure 4. As mentioned before, the latencies are presented in logarithmic scale. For most systems, the read latencies are fairly stable, while they differ strongly in the actual value. Again, Cassandra, HBase, and Voldemort illustrate a similar pattern – the latency increases slightly for two nodes and then stays constant. Project Voldemort has the lowest latency of 230 $\mu s$ for one node and 260 $\mu s$ for 12 nodes. Cassandra has a higher average latency of 5 - 8 ms and HBase has a much higher latency of 50 - 90 ms. Both sharded stores, Redis, and MySQL, have a similar pattern as well, with Redis having the best latency among all systems. In contrast to the web data stores, they have a latency that tends to decrease with the scale of the system. This is due to the reduced load per system that reduces the latency as will be further discussed in Section 5.6. The latency for reads in VoltDB is increasing which is consistent with the decreasing throughput. The read latency is surprisingly high also for the single node case which, however, has a solid throughput.

---

[7]We tried both supported hashing algorithms in Jedis, MurMurHash and MD5, with the same result. The presented results are achieved with MurMurHash

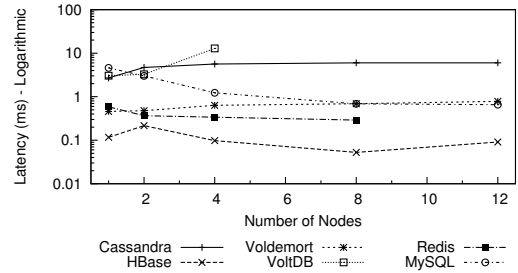

Figure 5: Write latency for Workload R

The latencies for write operations in Workload R can be seen in Figure 5. The differences in the write latencies are slightly bigger than the differences in the read latencies. The best latency has HBase which clearly trades a read latency for write latency. It is, however, not as stable as the latencies of the other systems. Cassandra has the highest (stable) write latency of the benchmarked systems, which is surprising since it was explicitly built for high insertion rates [19]. Project Voldemort has roughly the same write as read latency and, thus, is a good compromise for write and read speed in this type of workload. The sharded solutions, Redis and MySQL, exhibit the same behavior as for read operations. However, Redis has much lower latency then MySQL while it has less throughput for more than 4 nodes. VoltDB again has a high latency from the start which gets prohibitive for more than 4 nodes.

## 5.2 Workload RW

In our second experiment, we ran Workload RW which has 50% writes. This is commonly classified as a very high write rate. In Figure 6, the throughput of the systems is shown. For a single node, VoltDB achieves the highest throughput, which is only slightly lower than its throughput for Workload R. Redis has a similar throughput, but it has 20% less throughput than for Workload R. Cassandra

1729

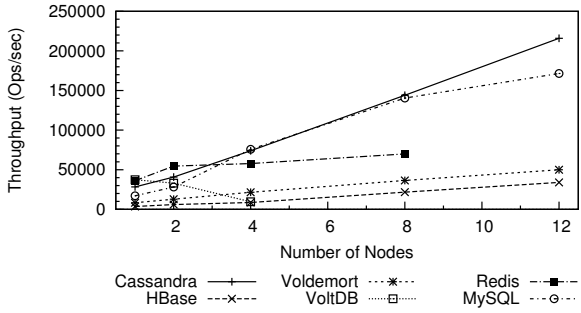

Figure 6: Throughput for Workload RW

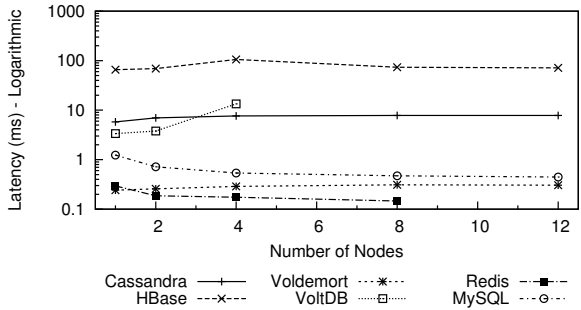

Figure 7: Read latency for Workload RW

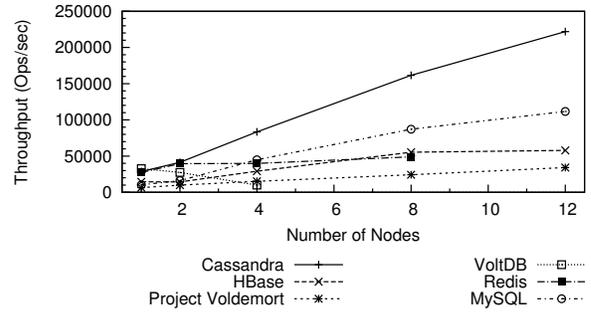

Figure 9: Throughput for Workload W

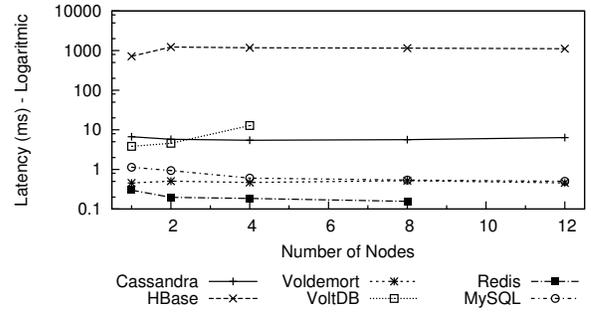

Figure 10: Read latency for Workload W

has a throughput that is about 10% higher than for the first workload. HBase's throughput increases by 40% for the higher write rate, while Project Voldemort's throughput shrinks by 33% as does MySQL's throughput.

For multiple nodes, Cassandra, HBase, and Project Voldemort follow the same linear behavior as well. MySQL exhibits a good speed-up up to 8 nodes, in which MySQL's throughput matches Cassandra's throughput. For 12 nodes, its throughput does no longer grow noticeably. Finally, Redis and VoltDB exhibit the same behavior as for the Workload R.

As can be seen in Figure 7, the read latency of all systems is essentially the same for both Workloads R and RW. The only notable difference is MySQL, which is 75% less for one node and 40% less for 12 nodes.

In Figure 8, the write latency for Workload RW is summarized. The trends closely follows the write latency of Workload R. However, there are two important subtle differences: (1) HBase's latency is almost 50% lower than for Workload R; and (2) MySQL's latency is twice as high on average for all scales.

## 5.3 Workload W

Workload W is the one that is closest to the APM use case (without scans). It has a write rate of 99% which is too high for web information systems' production workloads. Therefore, this is a workload neither of the systems was specifically designed for. The throughput results can be seen in Figure 9. The results for one node are similar to the results for Workload RW with the difference that all system have a worse throughput except for Cassandra and HBase. While Cassandra's throughput increases modestly (2% for 12 nodes), HBase's throughput increases almost by a factor of 2 (for 12 nodes).

For the read latency in Workload W, shown in Figure 7, the most apparent change is the high latency of HBase. For 12 nodes, it goes up to 1 second on average. Furthermore, Voldemort's read latency almost twice as high while it was constant for Workload R and RW. For the other systems the read latency does not change significantly.

The write latency for Workload W is captured in Figure 11. It can be seen that HBase's write latency increased significantly, by a factor of 20. In contrast to the read latency, Project Voldemort's write latency is almost identical to workload RW. For the other systems the write latency increased in the order of 5-15%.

## 5.4 Workload RS

In the second part of our experiments, we also introduce scans in the workloads. In particular, we used the existing YCSB client for Project Voldemort which does not support scans. Therefore, we omitted Project Voldemort in the following experiments. In the scan experiments, we split the read percentage in equal sized scan and read parts. For Workload RS this results in 47% read and scan operations and 6% write operations.

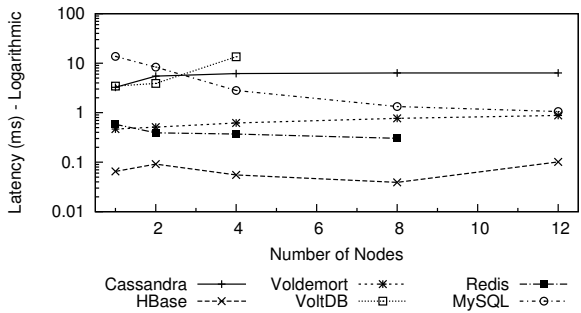

Figure 8: Write latency for Workload RW

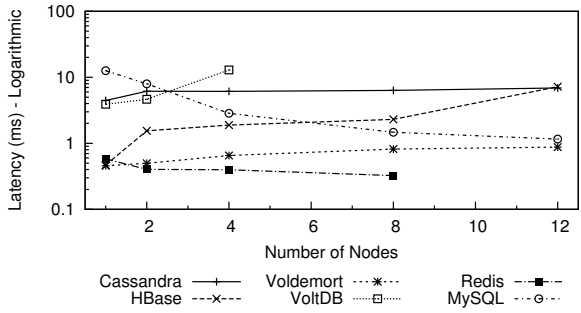

Figure 11: Write latency for Workload W

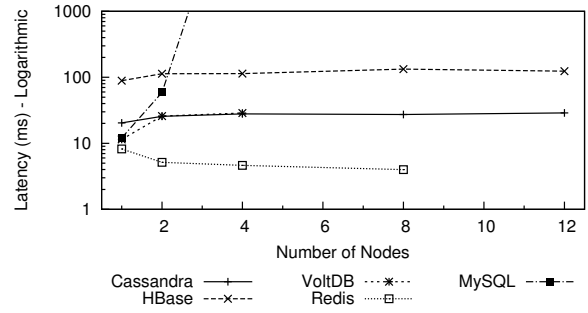

Figure 13: Scan latency for Workload RS

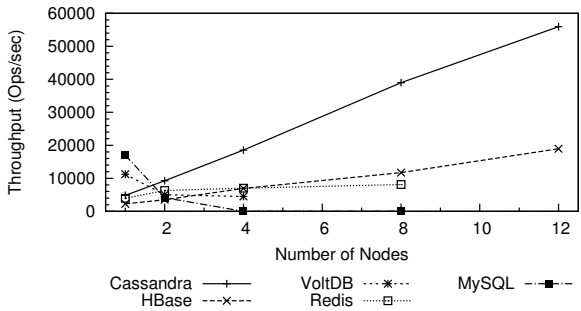

Figure 12: Throughput for Workload RS

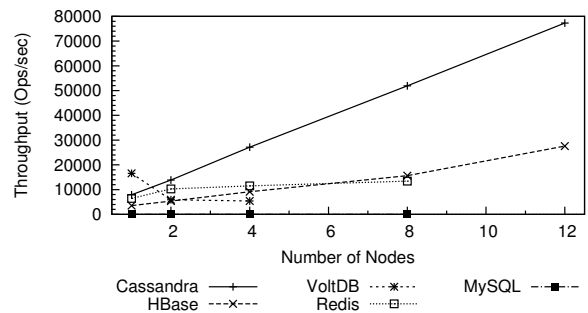

Figure 14: Throughput for Workload RSW

In Figure 12, the throughput results can be seen. MySQL has the best throughput for a single node, but does not scale with the number of nodes. The same is true for VoltDB which is, however, consistent with the general performance of VoltDB in our evaluation. Furthermore, Redis achieves similar performance as in the workloads without scans. Lastly, Cassandra and HBase, again, obtain a linear increase in throughput with the number of nodes.

The scan latency results, shown in Figure 13, signify that the MySQL scans are slow for setup with larger than two nodes. This justifies the low throughput. MySQL's weak performance for scans can also be justified with the way the scans are done in the YCSB client. The scan is translated to a SQL query that retrieves all records with a key equal or greater than the start key of the scan. In the case of MySQL this is inefficient.

HBase's latency is almost in the second range. Likewise, Cassandra's scans are constant and are in the range of 20-25 milliseconds. Although not shown, it is interesting to note that HBase's latency for read and write operations is the same as without scans. In addition, the scans are not significantly slower than read operations. This is not true for Cassandra, here all operations have the same increase of latency, and, in general, scans are 4 times slower than reads. Redis behaves like HBase, but has a latency that is in the range of 4-8 milliseconds. Similar to Cassandra, VoltDB has the same latency for all different operations.

### 5.5 Workload RSW

Workload RSW has 50% reads of which 25% are scans. The throughput results can be seen in Figure 14. The results are similar to workload RS with the difference that MySQL's throughput is as low as 20 operations per second for one node and goes below one operation per second for four and more nodes. This can again be explained with the implementation of the YCSB client. HBase and Cassandra gain from the lower scan rate and have, therefore, a throughput that is twice as high as for Workload RS. VoltDB achieves the best throughput for one node. Furthermore, VoltDB's throughput only slightly decreases from two to four nodes which can also be seen for Workload RS. The scan operation latency is for Workload RSW for Cassandra and Voldemort, all other systems have a slightly increased scan latency. The scan latencies are all stable with the exception of MySQL. We omit the graph due to space restrictions.

### 5.6 Varying Throughput

The maximum throughput tests above are a corner case for some of the systems as the high latencies show. To get more insights into the latencies in less loaded system we made a series of tests where we limited the maximum workload. In this test, we used the configuration with 8 nodes for each system and limited the load to 95% to 50% of the maximum throughput that was determined in the previous tests[8]. Due to space restrictions, we only present our results for Workload R.

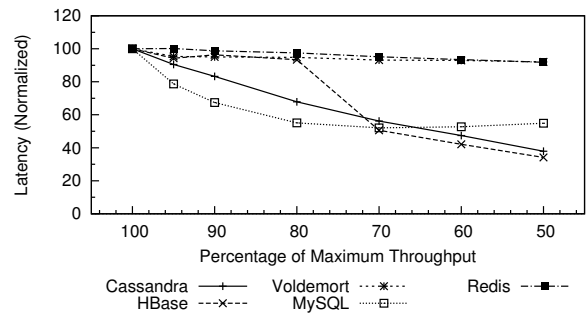

Figure 15: Read latency for bounded throughput on Workload R

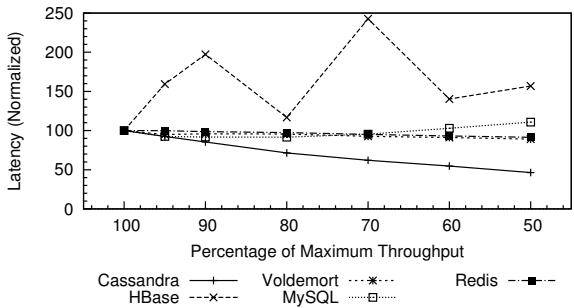

**Figure 16: Write latency for bounded throughput on Workload R**

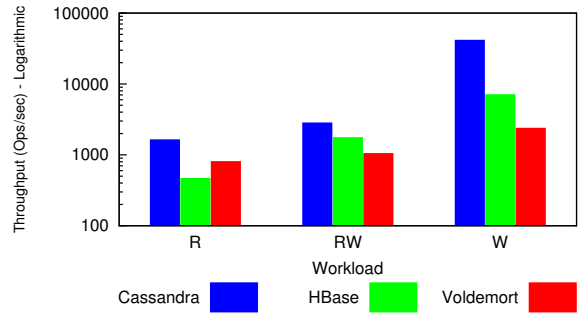

**Figure 18: Throughput for 8 nodes in Cluster D**

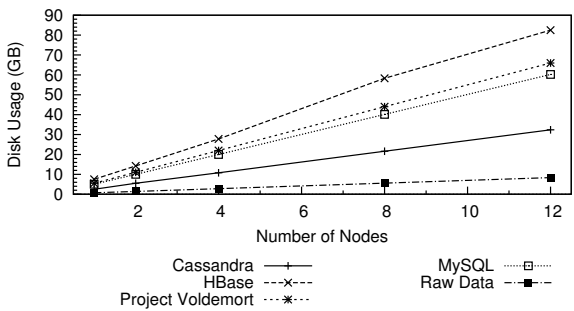

**Figure 17: Disk usage for 10 million records**

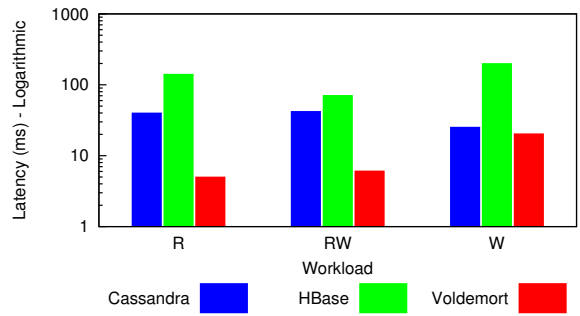

**Figure 19: Read latency for 8 nodes in Cluster D**

In Figure 15, the normalized read latency for Workload R can be seen. For Cassandra the latency decreases almost linearly with the reduction of the workload. Redis and Voldemort have only small but steady reductions in the latencies. This shows that for these systems the bottleneck was probably not the query processing itself. HBase has an interesting behavior that lets assume that the system has different states of operation based on the system load. Below 80% of the maximum load the read latency decreases linearly while being very constant above. For MySQL the latency first decreases rapidly and then stays steady which is due to the imbalanced load on the nodes.

The write latencies have a similar development for Cassandra, Voldemort, and Redis, as can be seen in Figure 16. HBase is very unstable, however, the actual value for the write latency is always well below 0.1 milliseconds. MySQL has a more constant latency as for the read latency.

### 5.7 Disk Usage

The disk usage of the key value stores initially came as a surprise to us. Figure 17 summarizes the disk usage of all systems that rely on disks. Since Redis and VoltDB do not store the data on disk, we omit these systems. As explained above, the size of each record is 75 bytes. Since we insert 10 million records per node, the data set grows linearly from 700 megabytes for one node to 8.4 gigabytes for 12 nodes. As expected, all system undergo a linear increase of the disk usage since we use no replication.

Cassandra stores the data most efficiently and uses 2.5 gigabytes per node after the load phase. MySQL uses 5 gigabytes per node and Project Voldemort 5.5 gigabytes. The most inefficient system in terms of storage is HBase that uses 7.5 gigabytes per node and therefore 10 times as much as the raw data size. In the case of MySQL, the disk usage also includes the binary log without this feature the disk usage is essentially reduced by half.

---
[8]Due to the prohibitive latency of VoltDB above 4 nodes we omitted it in this test.

The high increase of the disk usage compared to the raw data is due to the additional schema as well as version information that is stored with each key-value pair. This is necessary for the flexible schema support in these systems. The effect of increased disk usage is stronger in our tests then other setups because of the small records. The disk usage can be reduced by using compression which, however, will decrease the throughput and thus is not used in our tests.

### 5.8 Disk-bound Cluster (Cluster D)

We conducted a second series of tests on Cluster D. In this case, all systems had to use disk since the inserted data set was larger than the available memory. Therefore, we could not test Redis and VoltDB in this setup. We also omitted MySQL in this test, due to limited availability of the cluster. Also we only focused on a single scale for workloads R, RW, and W.

In Figure 18, the throughput on this system can be seen for all three workloads on a logarithmic scale. In this test, the throughput increases for all systems significantly with higher write ratios. This most significant result is for Cassandra which had relatively constant throughput for different tests in Cluster M. In Cluster D, the throughput increases by a factor of 26 from Workload R to Workload W. HBase's throughput also benefits significantly by factor of 15. Project Voldemort's throughput also increases only by a factor of 3. These results are especially interesting since these systems were originally designed for read-intensive workloads.

As can be seen in Figure 19, the read latencies of all systems are in the order of milliseconds. Cassandra has a read latency of

1732

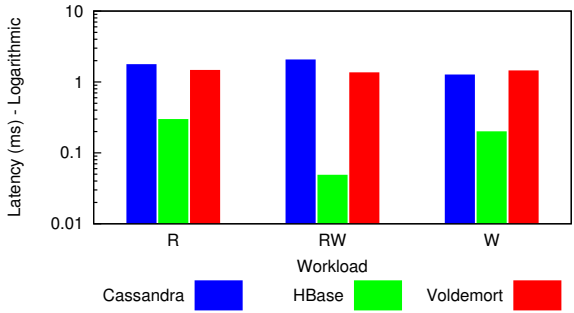

**Figure 20: Write latency for 8 nodes in Cluster D**

40 ms for Workload R and RW. For workload W the latency is 25 ms. HBase's read latency is surprisingly best in the mixed Workload RW with 70 ms on average, for Workload W it is worst with over 200 ms. Voldemort has by far the best latency that is 5 and 6 ms for Workload R and Workload RW and increases to 20 ms for Workload W.

The write latency is less dependent on the workload as can be seen in Figure 20. As in Cluster M, HBase has a very low latency, well below 1 ms. Interestingly, it is best for Workload RW. Cassandra and Project Voldemort exhibit a similar behavior. The write latency of these two systems is stable with a slight decrease for Workload RW.

## 5.9 Discussion

In terms of scalability, there is a clear winner throughout our experiments. Cassandra achieves the highest throughput for the maximum number of nodes in all experiments with a linear increasing throughput from 1 to 12 nodes. This comes at the price of a high write and read latencies. Cassandra's performance is best for high insertion rates. HBase's throughput in many experiments is the lowest for one node but also increases almost linearly with the number of nodes. HBase has a low write latency, especially in workloads with a considerable number of reads. The read latency, however, is much higher than in other systems. Project Voldemort in our tests positions itself in between HBase and Cassandra. It also exhibits a near linear scalability. The read and write latency in Project Voldemort are similar and are stable at a low-level. Cassandra, HBase, and Project Voldemort were also evaluated on Cluster D. In this disk-bound setup, all systems have much lower throughputs and higher latencies.

Our sharded MySQL installation achieves a high throughput as well which is almost as high as Cassandra's. Interestingly, the latency of the sharded system decreases with the number of nodes due to the decreased relative load of the individual systems. For scans, the performance of MySQL is low which is due to the implementation of scans in the sharding library. Since we were not able to successfully run the Redis cluster version, we used the sharding library Jedis. The standalone version of Redis has a high throughput that exceeds all other systems for read-intensive workloads. The sharding library, however, does not balance the workload well which is why the the throughput does not increase in the same manner as for MySQL. However, the latencies for both read and write operations also decrease with increasing number of nodes for the sharded Redis setup. Intrigued by the promised performance, we also included VoltDB in our experiments. The performance for a single instance is in fact high and comparable to Redis. However, we never achieved any throughput increase with more than one node.

## 6. EXPERIENCES

In this section, we report, from an industry perspective, additional findings and observations that we encountered while benchmarking the various systems. We report on the difficulty to setup, to configure, and, most importantly, to tune these systems for an industry-scale evaluation. In our initial test runs, we ran every system with the default configuration, and then tried to improve the performance by changing various tuning parameters. We dedicated at least a week for configuring and tuning each system (concentrating on one system at a time) to get a fair comparison.

### 6.1 YCSB

The YCSB benchmark was intuitive to use and fit our needs precisely. In the first version of YCSB that we used to benchmark the system, we, however, had a problem with its scalability. Because of the high performance of some of the systems under test, YCSB was not able to fully saturate them with one client node assigned to four storage nodes. Partly due to a recent YCSB patch (during our experimental evaluation) and by decreasing the ratio of client nodes to store nodes up to 1:2, we were able to saturate all data stores.

**Cassandra** Cassandra's setup was relatively easy, since there are quick-start manuals available at the official website[9]. Since Cassandra is a symmetric system, i.e., all nodes are equal, a single setup for all nodes is virtually sufficient. There are only a few changes necessary in the configuration file to setup a cluster of Cassandra. In our tests, we had no major issues in setting up the cluster.

Like other key-value store systems, Cassandra employs consistent hashing for distributing the values across the nodes. In Cassandra, this is done by hashing the keys in the a range of $2^{127}$ values and dividing this range by certain tokens. The default configuration selects a random seed (token) for each node that determines its range of hashed keys. In our tests, this default behavior frequently resulted in a highly unbalanced workload. Therefore, we assigned an optimal set of tokens to the nodes after the installation and before the load. This resulted in an optimally balanced data placement it, however, requires that the number of nodes is known in advance. Otherwise, a costly repartitioning has to be done for achieving a balanced data load.

**HBase** The configuration and installation of HBase was more challenge than in the case of Cassandra. Since HBase uses HDFS, it also requires the installation and configuration of Hadoop. Furthermore, HBase is not symmetric and, hence, the placement of the different services has an impact on the performance as well. Since we focused on a setup with a maximum of 12 nodes, we did not assign the master node and jobtracker to separate nodes instead we deployed them with data nodes.

During the evaluations, we encountered several additional problems, in which the benchmark unexpectedly failed. The first issue that randomly interrupted the benchmarks was a suspected memory leak in the HBase client that was also documented in the Apache HBase Reference Guide[10]. Although we were able to fix this issue with specific memory settings for the Java Virtual Machine, determine the root cause was non-trivial and demanded extensive amount of debugging. Another configuration problem, which almost undetectable was due to an incorrect permission setting for the HDFS data directory, in which the actual errors in log file were misleading. HBase is strict in the permission settings for its directories and does not permit write access from other users on the data directory. Although the error is written to the data node log, the error returned to the client reports a missing master.

---
[9]Cassandra website: http://cassandra.apache.org/
[10]Apache HBase Reference Guide - http://hbase.apache.org/book.html

In contrast to the other systems in our test, the HBase benchmark runs frequently failed when there was no obvious issue. These failures were non-deterministic and usually resulted in a broken test run that had to be repeated many times.

**Redis** The installation of Redis was simple and the default configuration was used with no major problems. However, Redis requires additional work to be done on the YCSB client side, which was implemented only to work against a single-node server instance. Since the Redis cluster version is still in development, we implemented our own YCSB client using the sharding capabilities of the Java Jedis library. Thus, we spread out a set of independent single-node Redis instances among the client nodes responsible of the sharding.

Therefore, as each thread was required to manage a connection for each of the Redis server, the system got quickly saturated because of the number of connections. As a result, we were forced to use a smaller number of threads. Fortunately, smaller number of threads were enough to intensively saturate the systems.

**Project Voldemort** The configuration of Project Voldemort was easy for the most part. However, in contrast to the other systems, we had to create a separate configuration file for each node. One rather involved issue was tuning the configuration of the client. In the default setting, the client is able to use up to 10 threads and up to 50 connections. However, in our maximum throughput setup this limit was always reached. This triggered problems in the storage nodes because each node configured to have a fixed number of open connections – which is in the default configuration 100 – this limit is quickly reached in our tests (with only two YCSB client threads). Therefore, we had to adjust the number of server side threads and the number of threads per YCSB instances. Furthermore, we had to optimize the cache for the embedded BerkeleyDB so that Project Voldemort itself had enough memory to run. For inadequate settings the server was unreachable after the clients established their connections.

**MySQL** MySQL is a well-known and widely documented project, thus the installation and configuration of the system was smooth. In short, we just set InnoDB as the storage system and the size of the buffer pool accordingly to the size of the memory.

As we used the default RDBMS YCSB client, which automatically shards the data on a set of independent database servers and forces each client thread to manage a JDBC connection with each of the servers, we required to decrease the number of threads per client in order to not saturate the systems. An alternative approach would be to write a different YCSB client and use a MySQL cluster version.

**VoltDB** Our VoltDB configuration was mostly inspired by the VoltDB community documentation[11] that suggested the client implementation and configuration of the store. The VoltDB developers benchmarked the speed of VoltDB vs. Cassandra themselves with a similar configuration but only up to 3 nodes [14]. Unlike our results they achieved a speed-up with a fixed sized database. In contrast to our setup, their tests used asynchronous communication which seems to better fit VoltDB's execution model.

## 7. RELATED WORK

After many years in which general purpose relational database systems dominated not only the market but also academic research, there has been an advent of highly specialized data stores. Based on the key-value paradigm many different architectures where created. Today, all major companies in the area of social Web have deployed a key-value store: Google has BigTable, Facebook Cassandra, LinkedIn Project Voldemort, Yahoo! PNUTS [6], and Amazon Dynamo [8]. In our benchmarking effort, we compared six well-known, modern data stores, all of which are publicly available. In our comparison, we cover a broad range of architectures and chose systems that were shown or at least said to be performant. Other stores we considered for our experiments but excluded in order to present a more thorough comparison of the systems tested were: Riak[12], Hypertable[13], and Kyoto Cabinet[14]. A high-level overview of different existing systems can be found in [4].

As explained above, we used the YCSB benchmark for our evaluation [7]. This benchmark is fairly simple and has a broad user base. It also fits our needs for the APM use case. In the YCSB's publication, a comparison of PNUTS, Cassandra, HBase, and MySQL is shown. Interestingly enough, there is no other large scale comparison across so many different systems available. However, there are a multiple online one-on-one comparisons as well as scientific publications. We summarize the findings of other comparisons without claiming to be exhaustive. Hugh compared the performance of Cassandra and VoltDB in [14], in his setup VoltDB outperformed Cassandra for up to 3 nodes on a data set of 500K values. We see a similar result for Workload R, however, for 4 and more nodes the VoltDB performance drastically decreases in our setup. In [16], Jeong compared Cassandra, HBase, and MongoDB on a three node, triple replicated setup. In the test, the author inserted 50M 1KB records in the system and measured the write throughput, the read-only throughput, and a 1:1 read-write throughput which is similar to our Workload RW. The results show that Cassandara outperforms HBase with less difference than we observed in our tests. MongoDB is shown to be less performant, however, the authors note that they observed high latencies for HBase and Cassandra which is consistent with our results. Erdody compared the performance and latency for Project Voldemort and Cassandra in [10]. In the three node, triple replicated setup, 1.5KB and 15KB records are used which is much larger than our record size. In this setup, the performance difference of Project Voldemort and Cassandra is not as significant as in our setup.

Pirzadeh et al. used YCSB to evaluate the performance of scan operations in key value stores [23]. The authors compared the performance of Cassandra, HBase, and Project Voldemort with a focus on scan workloads to determine the scan characteristics of these systems. Patil et al. developed the YCSB++ benchmark [22]. In their tests, they compared HBase to their own techniques with up to 6 nodes. An interesting feature of the YCSB++ benchmark is the more enhanced monitoring using an extension of Ganglia. A comparison of Cassandra, HBase, and Riak discussing their elasticity was presented by Konstantinou et al. [18]. The authors also use the YCSB benchmark suite. In the presented results, HBase consistently outperforms Cassandra which could be the case because of the older version of Cassandra (0.7.0 beta vs. 1.0.0-rc2).

We are not aware of any other study that compares the performance of such a wide selection of systems on a scale of up to 12+ nodes. In contrast to previous work, our data set consists of small records which increases the impact of inefficient resource usage for memory, disk and network. Our performance numbers are in many cases consistent with previous findings but give a comprehensive comparison of all systems for a broad range of workloads.

---

[11]VoltDB Performance Guide - http://community.voltdb.com/docs/PerfGuide/index

[12]Riak homepage - http://wiki.basho.com/

[13]Hypertable homepage - http://hypertable.org

[14]Kyoto Cabinet homepage - http://fallabs.com/kyotocabinet/

## 8. CONCLUSION

In this paper, we present the challenge of storing monitoring data as generated by application performance management tools. Due to their superior scalability and high performance for other comparable workloads, we analyzed the suitability of six different storage systems for the APM use case. Our results are valid also for related use cases, such as on-line advertisement marketing, click stream storage, and power monitoring. Unlike previous work, we have focused on the maximum throughput that can be achieved by the systems. We observed linear scalability for Cassandra, HBase, and Project Voldemort in most of the tests. Cassandra's throughput dominated in all the tests, however, its latency was in all tests peculiarly high. Project Voldemort exhibits a stable latency that is much lower than Cassandra's latency. HBase had the least throughput of the three but exhibited a low write latency at the cost of a high read latency. The sharded systems, i.e., Redis and MySQL, showed good throughput that was, however, not as scalable as the first three systems' throughput. It has to be noted, however, that the throughput of multiple nodes in the sharded case depends largely on the sharding library. The last system in our test was VoltDB, a shared-nothing, in-memory database system. Although it exhibited a high throughput for a single node, the multi-node setup did not scale.

In our tests, we optimized each system for our workload and tested it with a number of open connections which was 4 times higher than the number of cores in the host CPUs. Higher numbers of connections led to congestion and slowed down the systems considerably while lower numbers did not fully utilize the systems. This configuration resulted in an average latency of the request processing that was much higher than in previously published performance measurements. Since our use case does not have the strict latency requirements that are common in on-line applications and similar environments, the latencies in most results are still adequate. Considering the initial statement that a maximum of 5% of the nodes are designated for storing monitoring data in a customer's data center, for 12 monitoring nodes, the number of nodes monitored would be around 240. If agents on each of these report 10K measurements every 10 seconds, the total number of inserts per second is 240K. This is higher than the maximum throughput that Cassandra achieves for Workload W on Cluster M but not drastically. However, since data is stored in-memory on Cluster M further improvements are needed in order to reliably sustain the requirements for APM.

In future work, we will determine the impact of replication and compression on the throughput in our use case. Furthermore, we will extend the range of tested architectures.

## 9. ACKNOWLEDGEMENTS

The authors would like to acknowledge the Invest in Spain society, the Ministerio de Economa y Competitividad of Spain and the EU through FEDER funds for their support through grant C2 10 18. Furthermore, we would like to thank Harald Kosch and his group for the generous access to his compute cluster.